\documentclass[a4paper,12pt,doublecol]{epl2_modified} %for 2 columns style without line numbers
\usepackage{amssymb} %DAS'S ADDITION%
\usepackage{color, soul} %DAS'S ADDITION%
\usepackage{amsmath} %DAS'S ADDITION%

\newcommand{\red}{\textcolor{black}}
\newcommand{\black}{\textcolor{black}}
\newcommand{\Q}{\mbox{WB}}

\newcommand{\QC}{\mbox{zeros}}
\usepackage{etoolbox} %JIM'S ADDITION%
\newtoggle{BRACKETS}
\toggletrue{BRACKETS}  %
\title{Measuring quasiperiodicity}
\shorttitle{Measuring quasiperiodicity} %Insert here a short version of the title if it exceeds 70 characters

\author{Suddhasattwa Das\inst{1} \and Chris B. Dock\inst{2} \and Yoshitaka Saiki\inst{3} \and Martin Salgado-Flores \inst{4} \and Evelyn Sander \inst{5} \and Jin Wu \inst{6} \and James A. Yorke \inst{7}}
\shortauthor{S. Das \etal}

\institute{                    
  \inst{1} Department of Mathematics, University of Maryland, College Park\\
  \inst{2} University of California, Berkeley \\
  \inst{3} Graduate School of Commerce and Management, Hitotsubashi University\\
  \inst{4} College of William and Mary\\
  \inst{5}  Department of Mathematical Sciences, George Mason University\\
  \inst{6} University of Maryland, College Park\\
  \inst{7} Department of Mathematics, Physics, and IPST, University of Maryland, College Park\\
}

\abstract{\red{
The Birkhoff Ergodic Theorem asserts under mild conditions that Birkhoff averages (i.e. time averages computed along a trajectory) converge to the space average. 
For sufficiently smooth systems, our small modification of numerical Birkhoff averages significantly speeds the convergence rate for quasiperiodic trajectories
-- by a factor of $10^{25}$ for 30-digit precision arithmetic, making it a useful computational tool for autonomous dynamical systems. 
}% red
Many dynamical systems and especially Hamiltonian systems are a complex mix of chaotic and quasiperiodic behaviors, and chaotic trajectories near quasiperiodic points can have long near-quasiperiodic transients. 
\red{
Our method can help determine which initial points are in a quasiperiodic set and which are chaotic.
We  use our {\bf weighted Birkhoff average} to study  quasiperiodic systems, 
to distinguishing between chaos and quasiperiodicity, 
and for computing rotation numbers for self-intersecting curves in the plane.}
\red{
Furthermore we introduce the Embedding Continuation Method which is a significantly simpler, general method for computing rotation numbers.
}
}

\begin{document}

\maketitle

%-_-_-_-_-_-_-_-_-_-_-_-_-_-_-_-_-_-_-_-_-_-_-_-_-_-_-_-_-_-_-_-_-_-_-_-_-_-_-_-_-_-_-_-_-_-_-_-_-_-_-_-_-_-_-_-_-_-_-_-_-_-_-_-_-_-_-_-_-_-_-_-_-_-_-_-
\section{Introduction}

Periodicity, quasiperiodicity, and chaos are the only three types of commonly observed dynamical behaviors in both deterministic models and experiments~\cite{sander:yorke:15}. 
A {\bf quasiperiodic orbit} of a map $T$ lies on a closed curve (or torus in higher dimensions) $X$, such that by a smooth change of coordinates, the dynamics of $T$ becomes pure rotation on the circle (resp. torus) by a fixed irrational {\bf rotation number(s)} $\rho$; 
\red{
that is, after the change in coordinates, the map on each coordinate $\theta_i$ becomes $\theta_i \mapsto \theta_i + \rho_i \mod1$ . }%red

Our improved method for computing Birkhoff averages for quasiperiodic trajectories enables the computation of rotation numbers, which are key parameters of these orbits. It also allows computation of the torus on which an orbit lies and of the change of coordinates that converts the dynamics to a pure rotation. 
\red{
%[[k]], 
%[[3]]
Our time series data is not appropriate for an FFT, but there is a standard way of computing such a change of coordinates using Newton's method to find Fourier series coefficients; 
see for example \cite{NumericQuasi5} by Jorba. It is effective if a few coefficients ($<100$) are needed. }%red

These quasiperiodic orbits occur in both Hamiltonian and more general systems~\cite{luque:villanueva:14,Simo1,Simo2,seara:villanueva:06,Durand:2002ug,Baesens:1990vj,Broer:1993uv,Vitolo:2011it,Hanssmann:2012jy,
Sevryuk:2012dv,Broer:1990ip,Kuznetsov:2015dl}. 
\black{
Luque and Villanueva~\cite{luque:villanueva:14} have published an effective method for computing rotation numbers, see their Figure 11.
On restricted three-body problems, they get 30-digit precision for rotation numbers using $N\approx 2 \times 10^6$ trajectory points while we get 30-digit precision with $N = 20,000$. In this paper, they apply their technique to rotation numbers and not other function integrals, 
but see also~\cite{luque:fourier}, where they used a slower convergence method for Fourier series.
}
More detail about our results here can be found in \cite{DSSY} (numerical) and \red{Corollary 2.1 from \cite{Das-Yorke} (theoretical).} 
\red{We should note that the Birkhoff approach (and ours) assumes we have a trajectory on the (quasiperiodic) set. If a quasiperiodic curve or torus is a hyperbolic set, that is, if it is unstable forwards and backwards in time, it might be quite a challenge to find a high precision trajectory on the curve.
}%red

Distinguishing between quasiperiodic and chaotic behavior in borderline cases is a difficult and important current topic of research for both models and experiments in physics~\cite{Zhusubaliyev15, Madhok15, Grebogi:83, Grebogi:85} and biology ~\cite{Ievlev14,Nicolis13}, and finding good numerical methods is a subject of active study~\cite{Sala15}. The coexistence of chaos and quasiperiodicity arbitrarily close to each other in a fractal pattern makes this detection a difficult problem. Recently proposed
methods~\cite{Levnajic-Mezic1, Levnajic-Mezic2} successfully distinguish between different invariant sets, but the methods suffer from extremely slow convergence due to their reliance on the use of Birkhoff averages. \red{By combining~\cite{Levnajic-Mezic1, Levnajic-Mezic2} with our method of weighted Birkhoff averages, we are able to distinguish between chaos and quasiperiodicity with excellent accuracy,} even in cases in which other methods of chaos detection such as the method of Lyapunov exponents, fail to give decisive answers. 

%-_-_-_-_-_-_-_-_-_-_-_-_-_-_-_-_-_-_-_-_-_-_-_-_-_-_-_-_-_-_-_-_-_-_-_-_-_-_-_-_-_-_-_-_-_-_-_-_-_-_-_-_-_-_-_-_-_-_-_-_-_-_-_-_-_-_-_-_-_-_-_-_-_-_-_-
\section{The Birkhoff average}
For a map $T$, let $x_n = T^n x$ be either a chaotic or a quasiperiodic trajectory. The Birkhoff average of a function $f$ along the trajectory is 
\begin{equation}
B_N(f)(x): = \Sigma_{n=0}^{N-1} (1/N)f(T^n(x)).
\end{equation}
Under mild hypotheses the Birkhoff Ergodic Theorem concludes that $B_N(f)(x) \to \int fd\mu$ as $N\to\infty$ where $\mu$ is an invariant probability measure for the trajectory's closure. This relationship between the time and space averages is incredibly powerful, allowing computation of $\int fd\mu$ whenever a time series is the only information available. However, the convergence of the Birkhoff average is slow, with an error of at least the order $N^{-1}$ for a length $N$ trajectory in the quasiperiodic case. 

\begin{center}
\begin{figure}
\includegraphics[width=.44\textwidth]{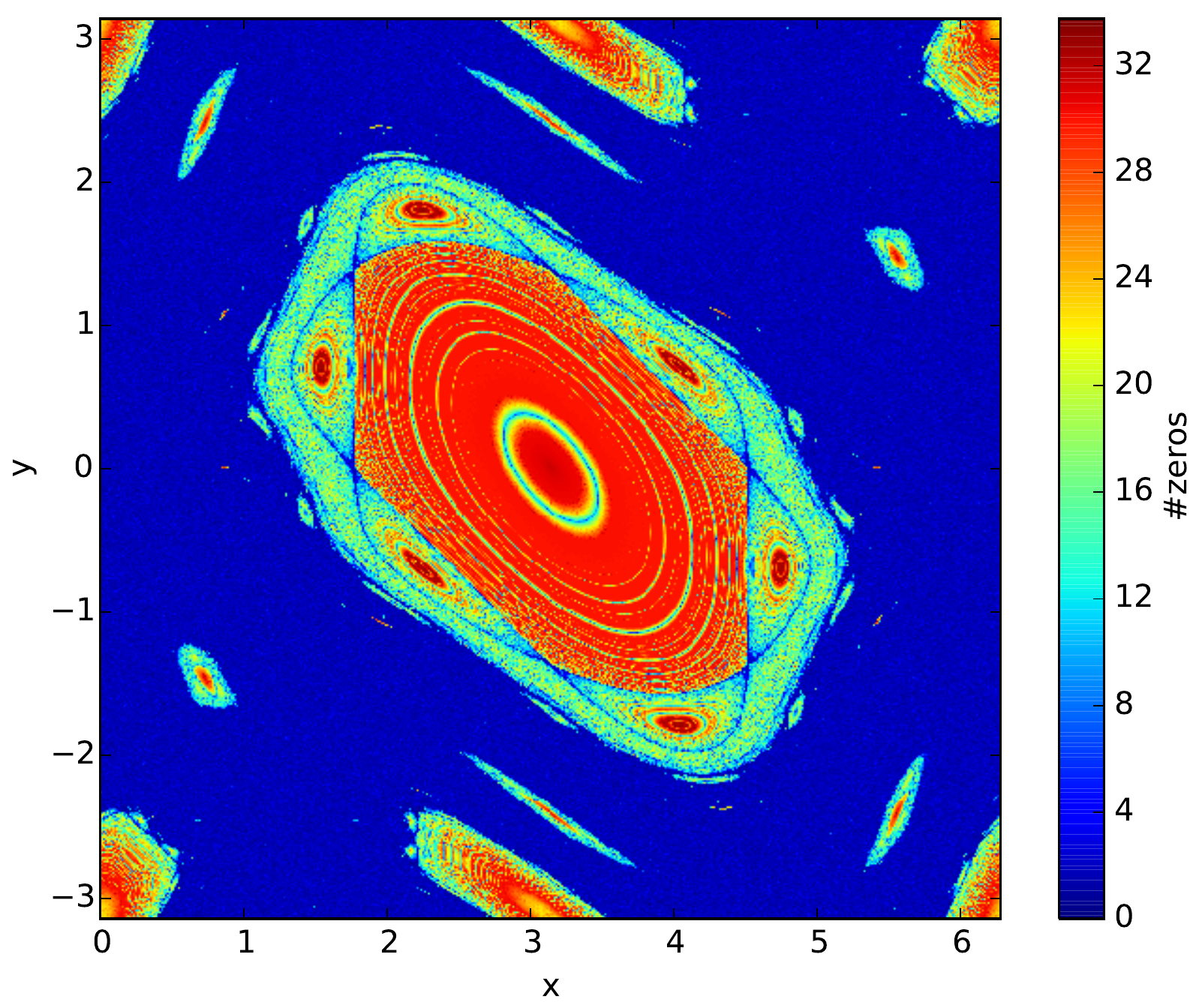}
\caption{\label{figchaos} {\bf Regions of chaos and quasiperiodicity for the Standard Map.} Here $r = 1.4$ and $\QC_N$ is calculated with $N=20,000$ and $f(x,y)=\sin(x+y)$. The value of $\QC_N$ is indicated by color coding. The dark blue region is chaotic, and all other colors indicate quasiperiodicity. Convergence of $\Q_N$ in the quasiperiodic region is slower (yellow to green) when the rotation number of an orbit is close to a rational number $m/n$ where $n$ is small such as $1/5$ or $1/6$. \red{See Corollary 2.1 in \cite{Das-Yorke} for details of the calculation. When $N$ is increased to $10^6$, 
almost all of the quasiperiodic points in the $500\times500$-point set displayed become red. }
}
\end{figure}
\begin{figure}
\includegraphics[width=.46\textwidth]{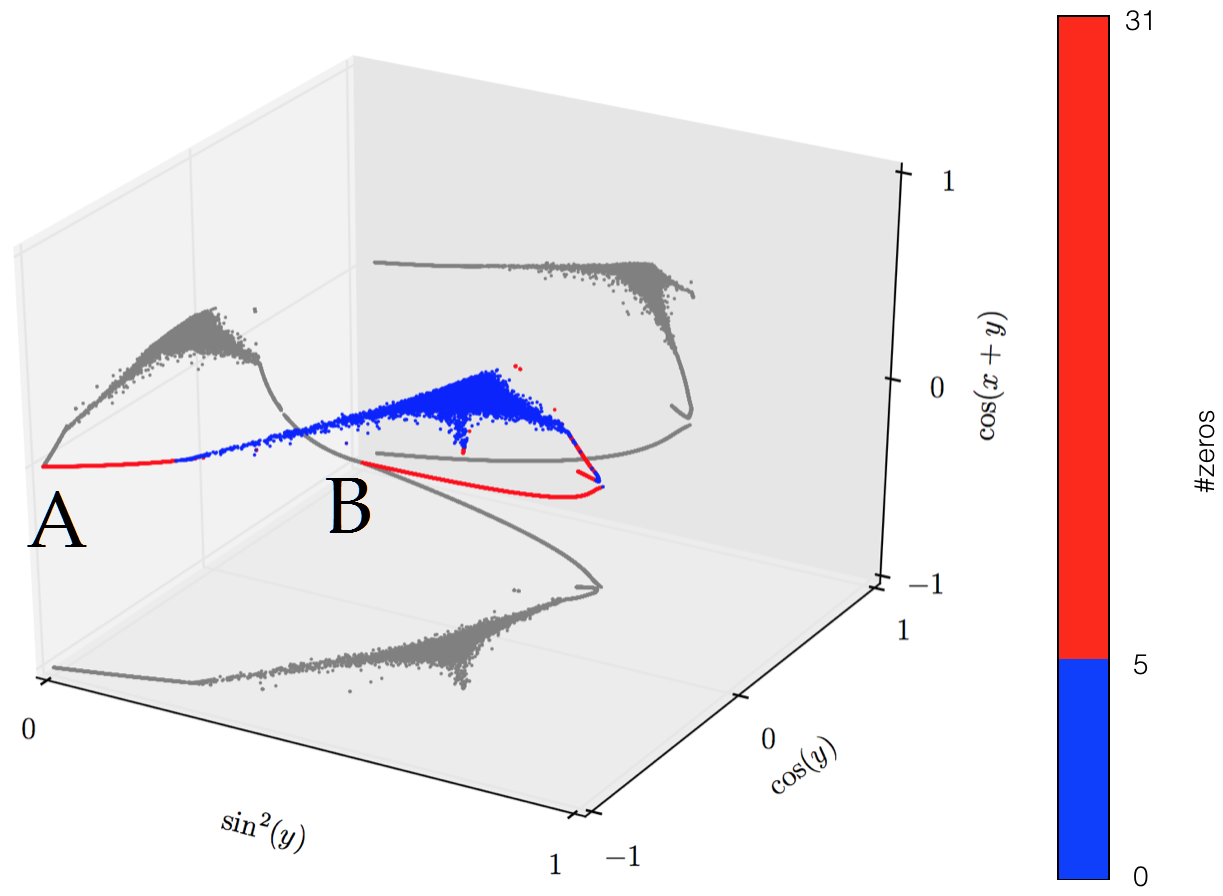}
\caption{\label{figembed} {\bf A three-dimensional embedding of the chaotic and quasiperiodic sets;} (as proposed in \cite{Levnajic-Mezic1, Levnajic-Mezic2}). 
For each initial condition in a grid on the torus, we set $N = 20,000$ and compute $\Q_N$ for the three different functions indicated on the axes.
Since $N=20,000$ is large enough to get excellent accuracy \emph{if} the point is quasiperiodic, all the points on a single orbit will yield the same (red) point in the plot. Hence a quasiperiodic disk yields a curve. 
Points in the chaotic region (blue) have considerable variation so the chaotic region results in a fuzzy shape.   
The gray sets are projections of the three-dimensional 
set onto the three coordinate planes.
The points A (front left side) and B (back left corner) correspond to the corner and center (respectively) of the torus in Fig 1.
}
\end{figure}
\begin{figure}
\includegraphics[width=.235\textwidth]{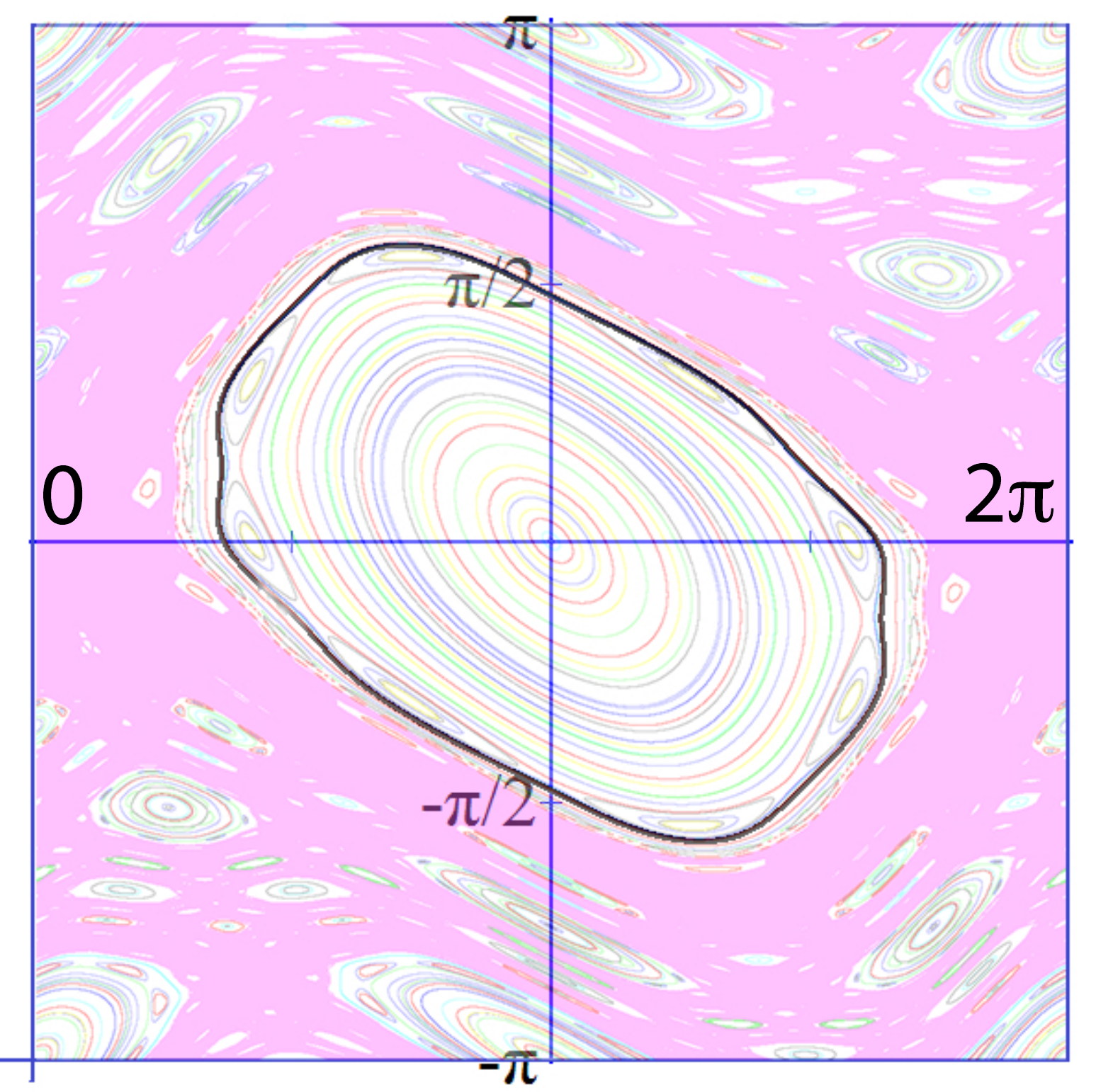}
\includegraphics[width=.235\textwidth]{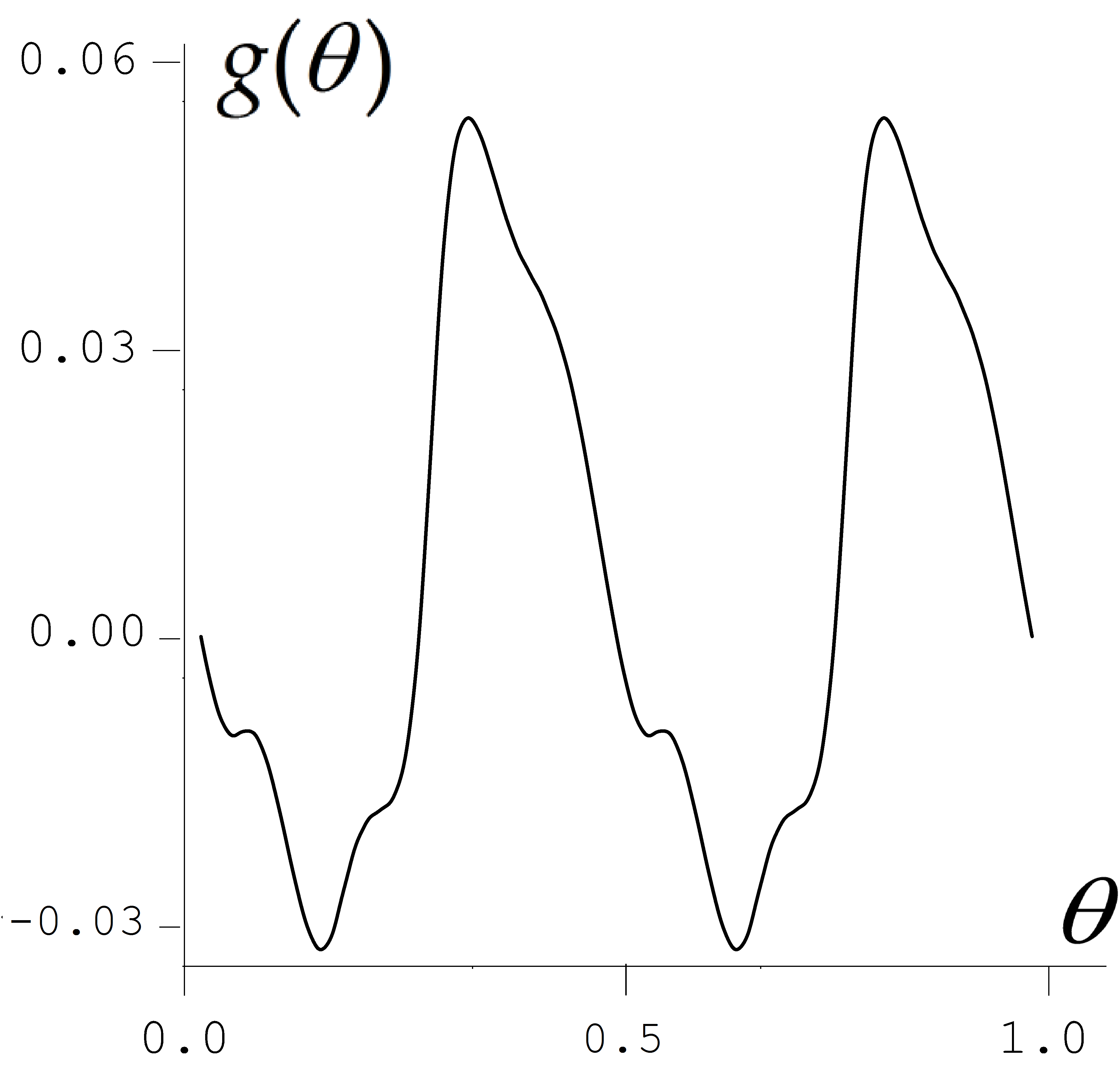}
\includegraphics[width=.235\textwidth]{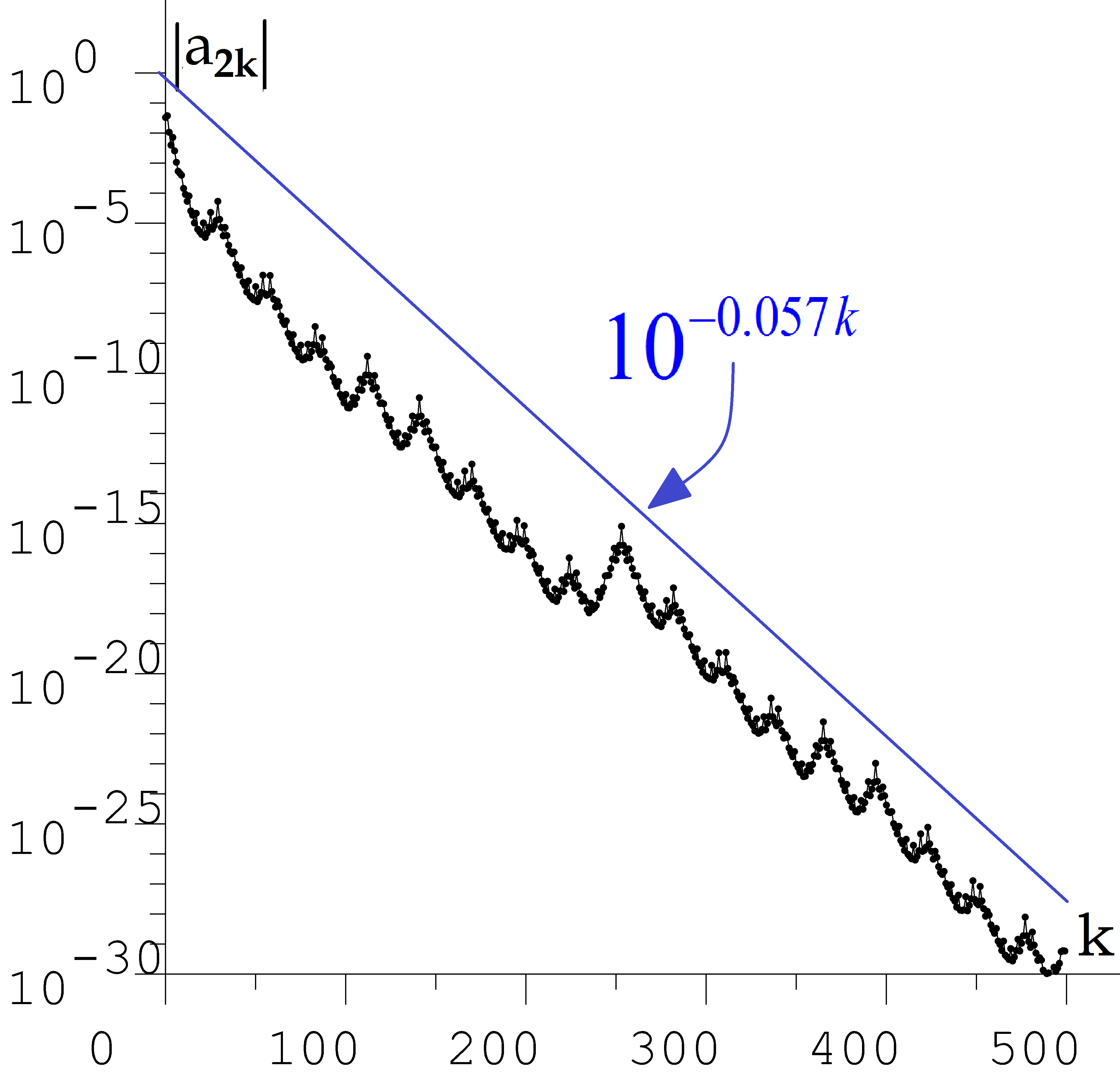}
\includegraphics[width=.235\textwidth]{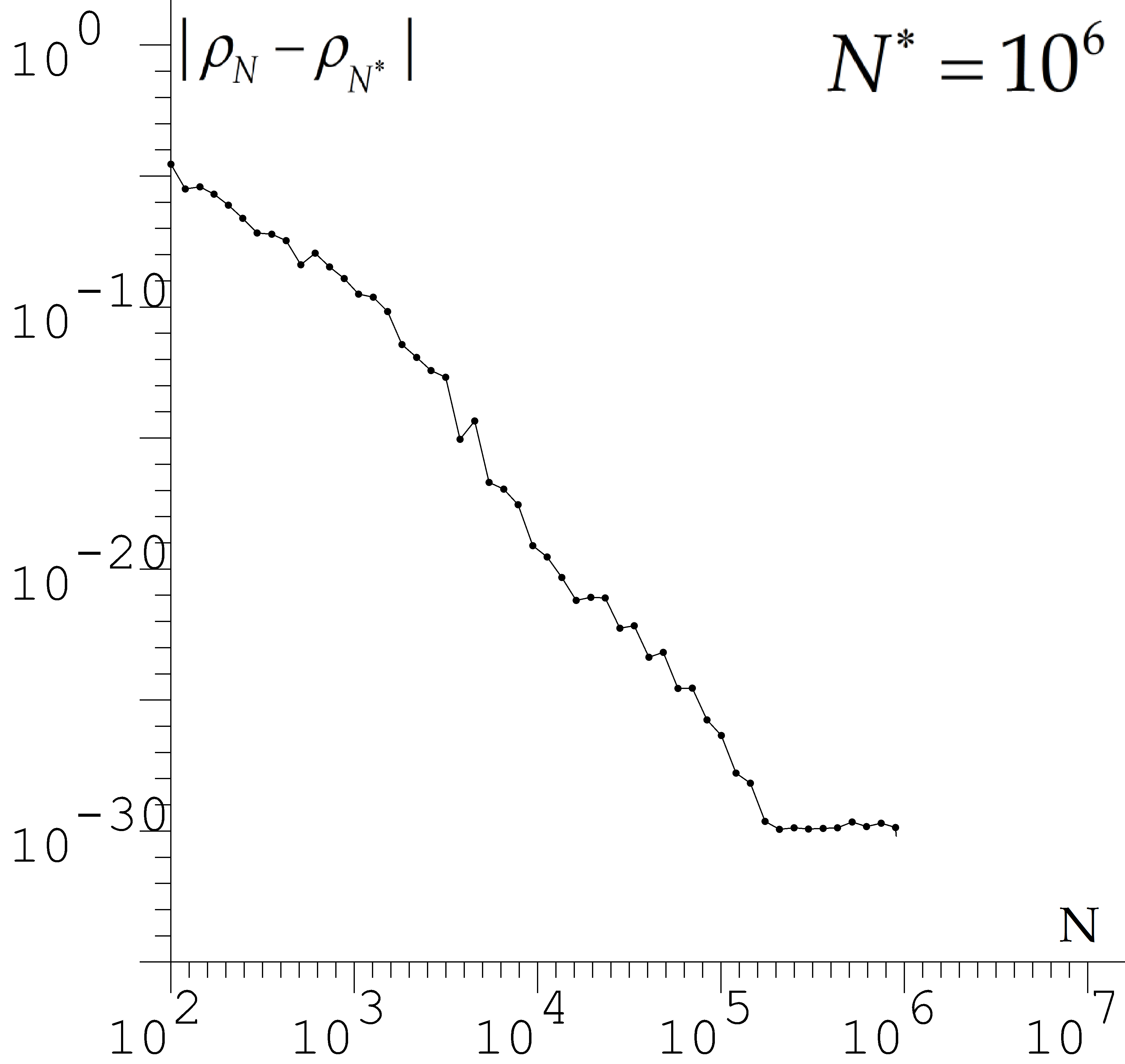} 
\caption{\textbf{A quasiperiodic circle for the Standard Map with $r=1.0$.}
Top left : The curve. Top right: The function $g(\theta) = \phi(\theta)-\theta$, the periodic part of the 
change of coordinates between the quasiperiodic circle and the pure rotation with rotation $\rho \approx 0.121$. Bottom left: The exponential decay of the Fourier coefficients of $g(\theta)$, only shown for even $k$ because $a_k=0$ for all odd $k$ . Note that $|a_{k}|=|a_{-k}|$. Bottom right: The super convergence of $\Q_N$ for the rotation number of Standard Map. \red{See Corollary 2.1 from }\cite{Das-Yorke} \red{for details of the calculation.} Note that the $|a_k|$ has a local maximum at $k=506$. 
\red{
%[[g]]
This type of spike can occur in the presence of small divisors, but we have verified that the spike correctly reflects the Fourier series.  See discussion.}
}
\label{figstandard2}
\end{figure}
\end{center}
{\bf Weighted Birkhoff ($\Q_N$) average. }
Instead of using Birkhoff's uniform weighting of $f(x_n)$, our average of these values gives very small weights to the terms $f(x_n)$ when $n$ is near $0$ or $N$.  
%[[l]]
Set
$w(t) := \red{exp(-[t(1-t)]^{-1})}$ for $t\in(0,1)$ and $=0$ elsewhere. 
Define the {\bf Weighted Birkhoff average ($\Q_N$)} of $f$ as follows. 
\begin{equation}\label{eqn:QN}
\Q_N(f)(x) :=\Sigma_{n=0}^{N-1} \hat{w}_{n,N}f(x_n),
\end{equation}
where $\hat{w}_{n,N}=w(n/N)/ \Sigma_{j=0}^{N-1}w(j/N)$. 
\red{
$\Q_N(f)$ has the same limit $\int_{}f$ as the Birkhoff average but on quasiperiodic trajectories $\Q_N(f)$ converges to that limit with 30-digit precision faster than $B_N$ {\bf by a factor of about \boldmath $10^{25}$ \unboldmath}.} %red
\red{
(There is no increase in convergence rate for chaotic trajectories). Intuitively, the improvement arises since the weight function $w$ vanishes at the ends, and thus gets rid of edge effects. 
We have proved  ~\cite{Das-Yorke} that if $(x_n)$ is a quasiperiodic trajectory and $f$ and $T$ are infinitely differentiable (i.e. $C^\infty$), then our method has ``super-convergence'' to $\int f d\mu$, {\em i.e.} 
 for each integer $m$ there is a constant $C_m$ for which 
$$|\Q_N(f)-\int fd\mu| \le C_m N^{-m}\mbox{ for all }N \ge 0.$$
}

\red{
The assumptions on $w$ are that it is infinitely differentiable; that $w(t)$ and all its derivatives are both $0$ at $t = 0$ and $t = 1$; and $\int_0^1 w(t)dt > 0$. }

\red{
%[[d]] 
A more general class of such $C^\infty$ weight functions for $p\ge 1$ is 
$w^{[p]}(t) := exp(-[t(1-t)]^{-p})$ for $t\in(0,1)$ and $=0$ elsewhere.  Our examples use $w= w^{[1]}$ here, but $w^{[2]}$ is even faster when requiring 30-digit precision. It is no faster when requiring 15-digit precision.  
%[[a]] 
The above constant $C_m$ depends on (i) $w(t)$ and its first $m$ derivatives; (ii) the function $f(t)$ ; and (iii) the rotation number(s) of the quasiperiodic trajectory or more precisely, the small divisors arising out of the rotation vector. We do not have a sharp estimate on the size of the term $C_m$.}

As a result of this speed, we are able to obtain high precision values for $\int f d\mu$ with a short trajectory and relatively low computational cost, largely independent of the choice of the $C^\infty$ function $f$. We get high accuracy results for rotation numbers and change of coordinates to a pure rotation for the Standard Map and the three-body problem. For a higher-dimensional example and further details, see~\cite{DSSY}. 

\red{
%[[2]] 
In creating $\Q_N$, we were motivated by ``apodization'' in optics (especially astronomy and photography), where diffraction that is caused by edge effects of lenses or mirrors can be greatly decreased. 
Our weighting method is reminiscent of both Hamming windows and Hann (or Hanning) filters for a Fourier transform on small windows (see for example, 
\cite{window_Fourier, NumericQuasi6, NumericQuasi7}.
The analogue of $w$ usually has only a couple derivatives $= 0$ at the end points $t=0,1$ and so convergence rate is only slightly better than the convergence rate of the standard Birkhoff method~\cite{DSSY}.}%red

\section{Testing for chaos} 
$\Q_N$ also provides a quantitative method of distinguishing quasiperiodic trajectories from chaotic trajectories. Along a trajectory $x_n$, we can compare the value of $\Q_N(f)$ along the first $N$ iterates with $\Q_N(f)$ along the second $N$ iterates, i.e. we consider $\Delta_N =\Q_N(f)(x)-\Q_N(f)(T^N(x))$. For a quasiperiodic orbit, we expect $|\Delta_N|$ to be very small. To measure how small $|\Delta_N|$ is, we can count the number of zeros after the decimal point by defining 
\begin{equation}
\QC_N(f)(x) = - \log_{10} |\Delta_N|.
\end{equation}
If the orbit is chaotic then $|\Delta_N|\sim N^{-1/2}$ or slower, $\QC_N$ is small. Whereas if it is quasiperiodic, both $\Q_Nf(x)$ and $\Q_N(f(T^N(x)))$ have super convergence to $\int fd\mu$ and so $\Delta_N$ has super convergence to $0$, implying that $\QC_N$ is large. For example, see Figure~\ref{figchaos}.
\red{
%[[e]]
%[[c2]] 
To check the accuracy of our method, we tested $12,086$ initial conditions on the diagonal $\{x=y\}$ for the Standard Map (Eqn. \ref{eqn:StdMap}).  We found that  $99.8$ per cent of the initial conditions for which $\QC_N>18$ for $N=20,000$ are in fact quasiperiodic (based on the criterion $\QC_N \ge 30$ for $N=10^8$).}

{\bf The (Taylor-Chirikov) Standard Map.} The Standard Map~\cite{StdMap1}
\begin{eqnarray}\label{eqn:StdMap}
S_1
\left(
\begin{array}{c}
x \\
y
\end{array}
\right)
=
\left(
\begin{array}{c}
x+y \\
y+ r \sin (x+y)
\end{array}
\right)
\pmod{2\pi}
\end{eqnarray}
is an area-preserving map on the two-dimensional torus in which both chaos and quasiperiodicity occur for a large set of parameter values. In order to distinguish the fine structure of regions of quasiperiodic versus chaotic behavior, we have used the $\QC_N$ test for chaos. Fig.~\ref{figchaos} shows the resulting distinct regions of chaos and quasiperiodicity. A further characterization of chaos versus quasiperiodicity is depicted in Fig.~\ref{figembed}, where $\QC_N$ is computed for three different functions. 
\black{All points on a quasiperiodic orbit map to the same red point in $\mathbb{R}^3$. The chaotic orbits (blue) remain spread out.}%red

{\bf Comparison with Lyapunov exponents.} Lyapunov exponents are a measure of the average stretching in each direction: a chaotic set will have a positive Lyapunov exponent, whereas
a quasiperiodic set of an area-preserving map has no average stretching
in any direction, so both of its Lyapunov exponents are zero.
The traditional numerical
calculation of Lyapunov exponents 
has a slow convergence rate similar that of $B_N$. 
This is compounded by the fact that chaotic
curves trapped between quasiperiodic rings are likely to have Lyapunov
exponents quite close to zero, mimicking the surrounding
quasiperiodicity. Hence a highly sensitive test is needed.
The use of $\QC_N$ is a
significant improvement compared to using Lyapunov exponents.
An alternative approach is to compute Lyapunov exponents using the
weighted Birkhoff average $\Q_N$. Then one would get the convergence rates of $\Q_N$; 
see \cite{DSSY}.

%-_-_-_-_-_-_-_-_-_-_-_-_-_-_-_-_-_-_-_-_-_-_-_-_-_-_-_-_-_-_-_-_-_-_-_-_-_-_-_-_-_-_-_-_-_-_-_-_-_-_-_-_-_-_-_-_-_-_-_-_-_-_-_-_-_-_-_-_-_-_-_-_-_-_-_-
\red{
{\bf Computing rotation numbers.}
%[[j]] 
 Assume there is a $C^\infty$ function $\gamma:S^1\to \mathbb{R}^2$ (such as the projection of a curve $\Gamma$ in higher dimensions) 
whose rotation number we want to know. 
The goal is to compute the (irrational) rotation number $\rho$ based only on knowledge of a trajectory $\gamma_n := \gamma(n\rho)$ where $\rho$ is unknown to the observer.}%red \red{

\red{
The rotation number of $\gamma$ is the rotation number of the original curve, 
which is independent of the projection into $\mathbb{R}^2$. 
Changing $\rho$ by an integer does not change $\gamma_n$, so
 it is only possible to determine $\rho\mod 1$.   
Writing $\hat\gamma(x) := \gamma(-x)$, we see that  $\gamma_n := \gamma(n\rho) = \hat\gamma(n(1-\rho))$.   
Therefore $\gamma_n$ has rotation number $\rho$ and $1-\rho$, depending on which map is used to define $\gamma_n$, 
so we cannot distinguish $\rho\mod 1$ from $-\rho\mod 1$ using only the trajectory. }

\red{       
For $P\in\mathbb{R}^2$ where $P\notin \gamma(S^1)$, define 
$\phi(x) \in [0,1]\mod1 $ so that $e^{i2\pi \phi(x)} = (\gamma(x)~-~P )/|| \gamma(x) - P ||$ as a point in $S^1$.}%red 

\red{
The {\bf winding number} of $\gamma$ around $P$ is 
$W(P) := \int_0^1 \phi'(x +s)ds$.
Let
\begin{equation*}
\hat\Delta(x): = \int_x^{x+\rho} \phi'(s)ds
\end{equation*}
where $\phi'(s)\in\mathbb{R}^1 $ is the derivative of $\phi$. }%red 

\red{Write $\hat\Delta_n := \hat\Delta(n\rho),$ where $1 \le n \le N$. By the Ergodic theorem 
\begin{eqnarray*} 
\lim_{N\to\infty}B_N(\hat\Delta_n) = \int_0^1 \hat\Delta_n(x) dx =  \int_0^1 \int_0^{\rho} \phi'(x+s)ds \; dx\\
 =\int_0^{\rho} \int_0^1  \phi'(x+s)dx \; ds = W(P)\int_0^{\rho} ds = W(P)\rho.
 \end{eqnarray*}
Thus the rotation number is the limit of $B_N(\hat\Delta_n)$, which equals $\rho$  if $W(P)=1$. Note that we cannot determine the sign of $W(P)$ from $\gamma_n$. }%red 

\red{ We still need to find $ \hat\Delta$: Write $\Delta(x) := (\phi(x+\rho)-\phi(x)) \mod 1\in S^1$. Then
$\hat\Delta(x)\in \mathbb{R}$ is a {\bf lift} of $\Delta(x)\in S^1$; i.e., they differ by an (unknown) integer 
$m(x) :=\hat\Delta(x) - \Delta(x)$. }%red 

\red{
{\bf Our Embedding Continuation Method.} We briefly outline our new method for computing rotation numbers of quasiperiodic curves, which extends to higher dimensions. 
Let $N$ be given; we imagine $N\sim 10^5$ or $10^6$. 
Choose $K\ge2$ and define the {\bf delay coordinate embedding} $\Gamma(x) := (\gamma(x), \gamma(x+\rho),...,\gamma(x+(K-1)\rho))\in\mathbb{R}^{2K}$. We will use the Euclidean norm on $\mathbb{R}^{2K}$. 
We also write $\Gamma_n := (\gamma_n, \gamma_{n+1},...,\gamma_{n+K-1})$.
}\red{
By the Whitney and Takens Embedding Theorems, for almost every smooth function $\gamma$ (in the sense of prevalence), the map $\Gamma:S^1\to\mathbb{R}^{2K}$ is an embedding. That is, there are no self intersections of $\Gamma(S^1)$; see \cite{Embedology} and references therein.
How $\Gamma(S^1)$ bends in $\mathbb{R}^{2K}$ is hard to envision, but there will be some $\delta_0>0$ such that for each $p \in \Gamma(S^1)$, the set of points in $\Gamma(S^1)$ that are within a distance 
$\delta_0$ of $p$ is a single arc. And for larger $K$, we expect the curve to be less wiggly.}%red 

\newcommand{\gd}{\bar\Delta^*}

\red{
{\bf Extending by Continuation}
We will define a function $\gd$ at all the points  $\Gamma_n$ so that $\gd_n$ differs from $\hat\Delta_n$ by a constant $k^*$.
Initially we define it at just one point.
Define $\gd_0 := \Delta_0$. Then $\hat\Delta_0 - \gd_0$ is an integer that we will call the afore mentioned $k^*$.  }%red 

\red{
When $\gd_n$ is defined at $\Gamma_n$, we can extend the definition to all points $\Gamma_m$ in a small neighborhood by continuity. 
Choose the integer $k_m$ for which $\Delta_m + k_m$ is close to $\gd_n$ and define $\gd_m$ to be $\Delta_m + k_m$. If $\gd_m$ was already defined, this will not change its value. }%red 

\red{
Continue extending until $\gd$ is defined on all of $\Gamma(S^1).$
It is continuous (and smooth) because $\bar\Delta$ is. The two differ by only by $k^*$.
Notice $B_N(\gd_n)-B_N(\hat\Delta_n)=k^*$, 
so $B_N(\gd_n)$ converges to $\rho \pm k^*$. We of course use $\Q_N.$}%red

\newcommand{\lift}{\mathcal{L}}

\begin{center}
\begin{figure}
\includegraphics[width=.24\textwidth]{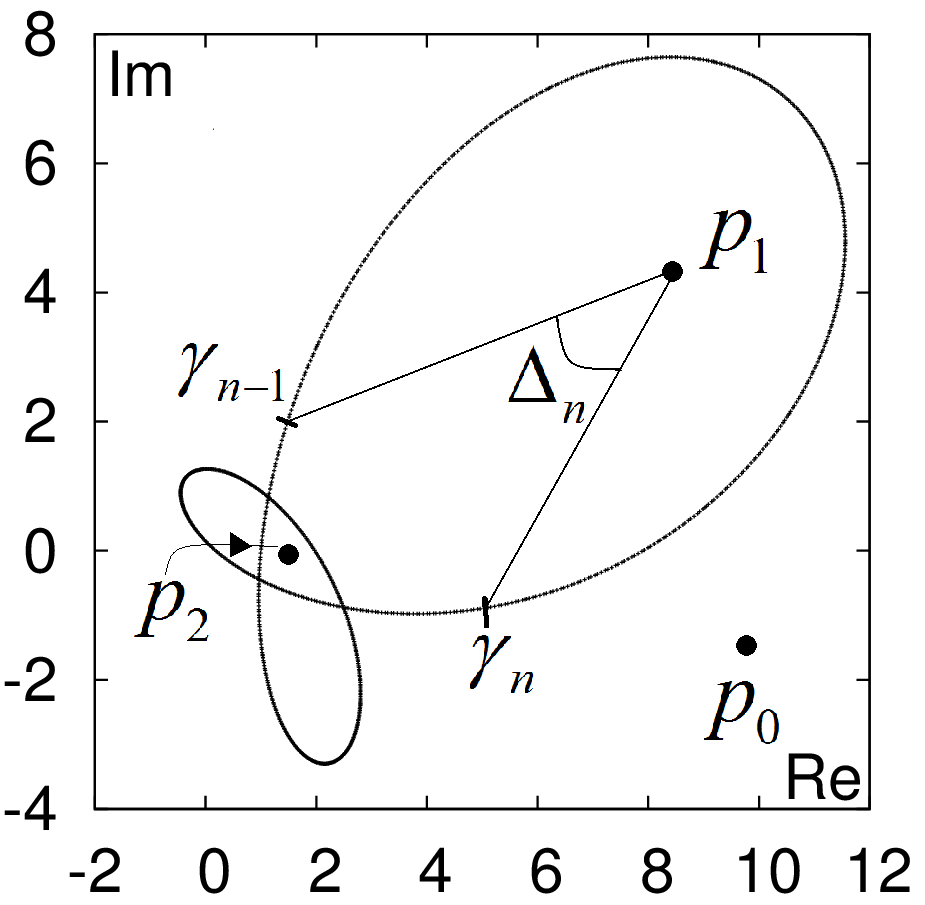}
\includegraphics[width=.24\textwidth]{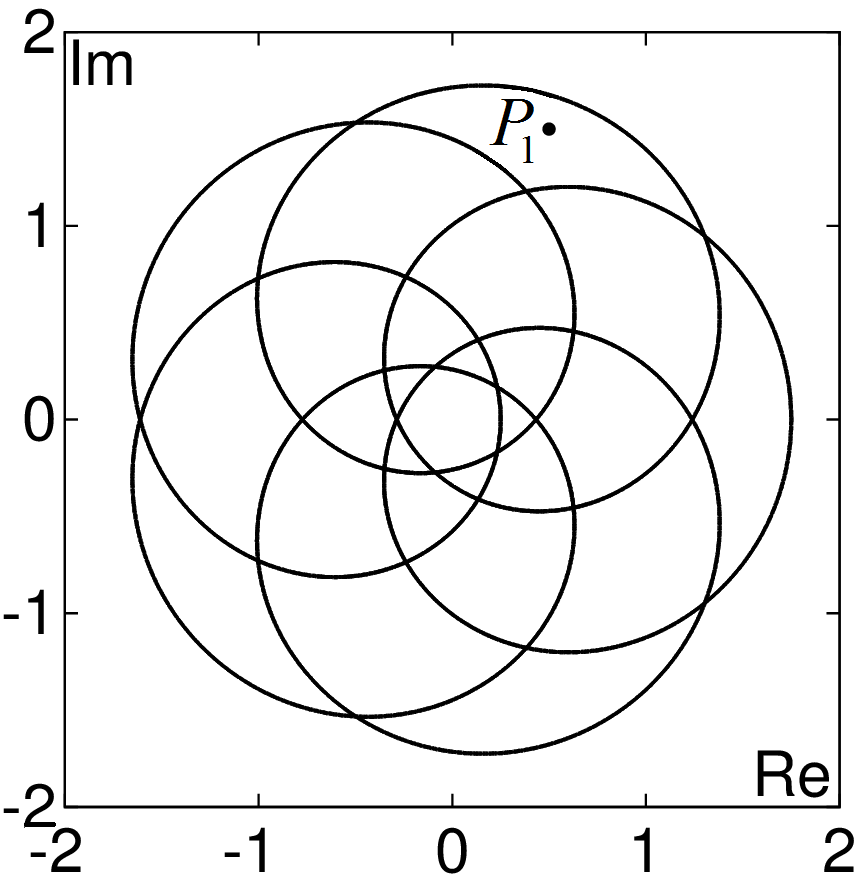}
\caption{\label{fig_Folded_Rot} \red{{\bf The folded curves ``the fish'' in Eq.~\ref{eqn:fish} and ``the flower'' in Eq.~\ref{eqn:flower}.}
The curves wind $j$ times around $P_j$
.}}
\end{figure}
\end{center}
\red{
{\bf Examples.}
Luque and Villanueva~\cite{ErgAvg9} addressed the case of a quasiperiodic planar curve $\gamma:S^1\to \mathbb{C}$ and introduced what we call ``the fish''. See Fig. \ref{fig_Folded_Rot} Left Panel.
\begin{equation}\label{eqn:fish}
  \gamma(x):=\hat{\gamma}_{-1}z^{-1}+\hat{\gamma_{0}}
  +\hat{\gamma_{1}}z+\hat{\gamma_2}z^2, 
\end{equation}
where $z = z(x):=e^{i 2\pi x}$  and  
$\hat\gamma_{-1}:=1.4-2i,$ $\hat\gamma_{0}:=4.1+1.34i,$ 
$\hat\gamma_{1}:=-2+2.412i,$ 
$\hat{\gamma_2}:=-2.5-1.752i$. 
(See Fig.5 and eq.(31) in~\cite{ErgAvg9}). 
They chose the rotation number $\rho=(\sqrt{5}-1)/2$ so that the trajectory is $\gamma_n = \gamma(n\rho)$ for $n = 0,1,\cdots$. The method in \cite{ErgAvg9} requires a step of  ``unfolding'' $\gamma$, which our method bypasses. We chose $P = P_1 = 7+4i$, where $|W(P_1)| = 1$.
We applied our embedding continuation method to define the $\gd_n$. Define $\rho_N := \Q_N(\gd_n)$. Fluctuations in $\rho_N$ fall below $10^{-30}$ for $N>20,000$. Since we know the actual rotation number, we can report that the error $|\rho-\rho_N|$ is  then below $10^{-30}$.}%red 

\red{
We call our next example ``the flower'',
Fig. \ref{fig_Folded_Rot} Right Panel. Let
\begin{equation}\label{eqn:flower}
\gamma_6(x): = (3/4) z + z^6 \mbox{ where } z = z(x) := e^{i 2 \pi x}.
\end{equation}
We chose $P=P_1:= (0.5, 1.5)$ for which $|W(P_1)|= 1$. We use the same $\rho$ as above.
There are points $P_j$ with $|W(P_j)| = j $ for $j= 0$ through $6$, and the origin $0$ has $W(0) = 6$ and for our method it is essential to choose a point $P$ where $|W(P)|=1.$
We note that $\max_x \hat\Delta(x) - \min_x \hat\Delta(x) \approx 1.2$.
The Embedding Continuation Method applied to Flower yields $\rho$ with 30-digit precision at $N=200,000$.}%red 

\section{Changing coordinates, making the map into a pure rotation}
Given a quasiperiodic trajectory $x_n$ in phase-space $M$, we are able
to construct a function $h:S^1\to M$, where $S^1$ is the unit circle, so
that $h(S^1)$ is the quasiperiodic curve on which the $x_n$ lie. 
\red{ We 
can use the methods described above to find a rotation number $\rho$ of a planar
quasiperiodic curve, 
such as the trajectory
of the Standard Map whose image is the black curve in the top-left panel
of Fig.~\ref{figstandard2}. 
}

We represent points in polar coordinates about the center of the domain so $h(\theta) =
(\phi(\theta),r(\theta))$, and we can
write $\phi(\theta)= \theta + g(\theta)$ where $g(\theta)$ is a bounded
periodic function. 

\red{
%[[g]] 
Once we know  the rotation number $\rho,$ we can determine the Fourier series for $g(\theta)=\Sigma_j a_j \sigma_j$, where $\sigma_j(\theta):=\exp(i2\pi j\theta)$ for each integer $j$. Note that $\sigma_0\equiv 1$.}

\red{Each Fourier coefficient of $g(\theta)$  is
%$\int_0^1 g(\theta)\exp(-2 i \pi k \theta)d\theta
$a_k:=\int_0^1 g(\theta)\sigma_{-k}(\theta)d\theta$. We compute 
$\hat{a}_k$ $:=\Q_N(g(
 \theta)\sigma_{-k}(\theta))$ to approximate $a_k$; substituting  $\Sigma_i a_i \sigma_i$ for $g(\theta)$ gives,
$$\hat{a}_k=\Sigma_i a_i \Q_N(\sigma_{i-k})=\Sigma_j a_{j+k} \Q_N(\sigma_{j}).$$
We find that for the Standard Map (Fig. \ref{figstandard2}), $\left|\Q_N(\sigma_j)\right|<10^{-32}$ when $N=4,000,000$ and $0<|j|\leq 1000$  . Therefore, the error in our estimate is
$$\left|\hat{a}_k-a_k\right|=\ \left|\Sigma_{j\neq 0}a_{j+k} \Q_N(\sigma_{j})\right|\ \approx 10^{-32}.$$
We construct $g(\theta)$ and therefore $h(\theta)=g(\theta)+\theta$ as a Fourier series and discover that  Fourier coefficients $a_k\to 0$ exponentially fast implying that  it is real analytic (at least to 30-digit precision). }

{\bf How smooth is the typical quasiperiodic curve?} 
First we point out that Yamaguchi and Tanikawa~\cite{StdMap1} and Chow et. al. \cite{1halfDegree} show that the outermost limit curve of the quasiperiodic sets in the Standard Map is not differentiable. But we have chosen a typical curve, not the most extreme. To answer this question for our curve, we examine its Fourier series. We find that the size of the Fourier coefficients decays exponentially fast, with size $ <10^{-30}$ by
the 500$^{th}$ coefficient. See Fig.~\ref{figstandard2}, bottom-left. This exponential decay rate, $|a_k| \le 10^{- const~k}$, is a characteristic of real-analytic functions, and we can therefore assert that the change of coordinates (both $\phi(\theta)$ and $g(\theta)$) are real-analytic (up to the quadruple precision of our calculation).
\red{Our calculation of the Fourier coefficients were based on an orbit of length $10^6$. To test the accuracy of our computations, these coefficients were used to predict the trajectory point after $10^7$ iterates and the prediction was correct by at least 22 decimal digits.}
\begin{figure}
\includegraphics[width=.235\textwidth]{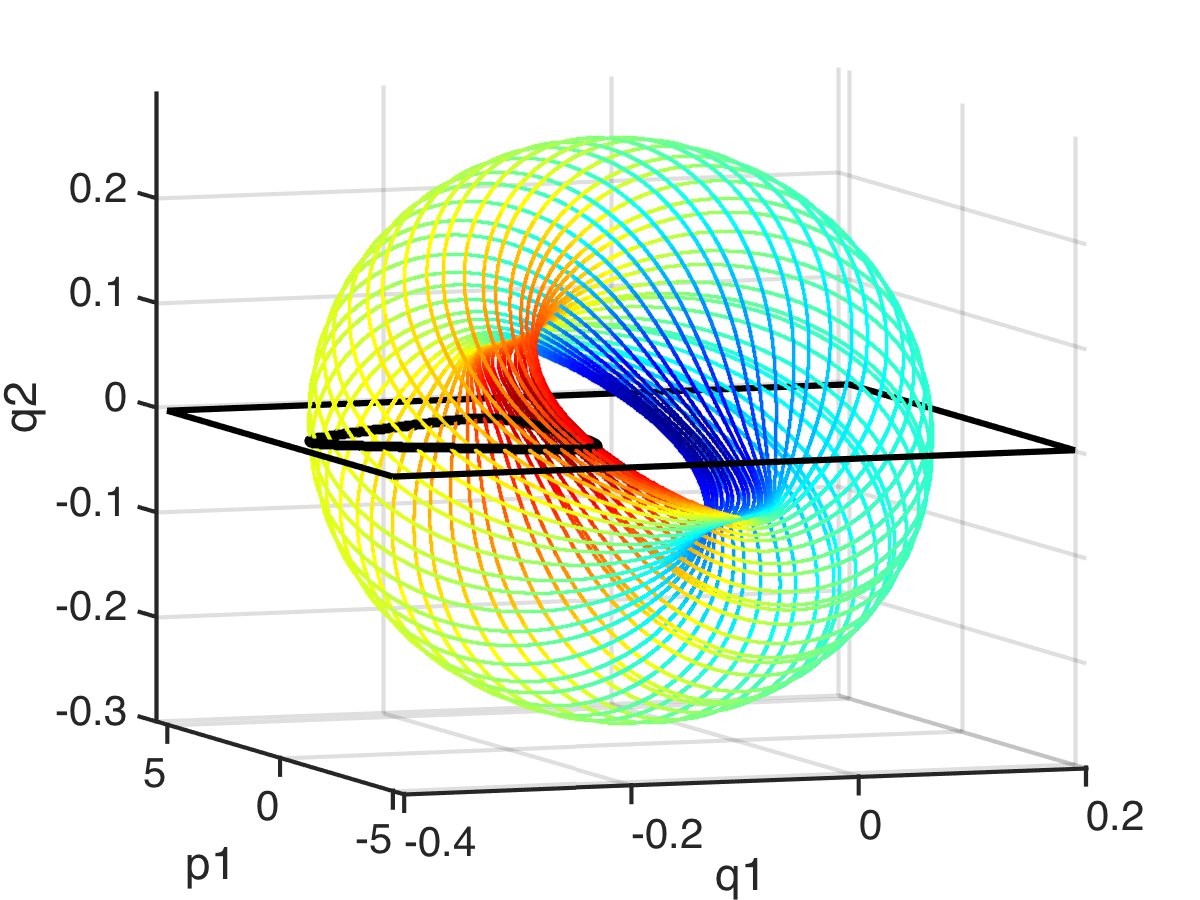}
\includegraphics[width=.235\textwidth]{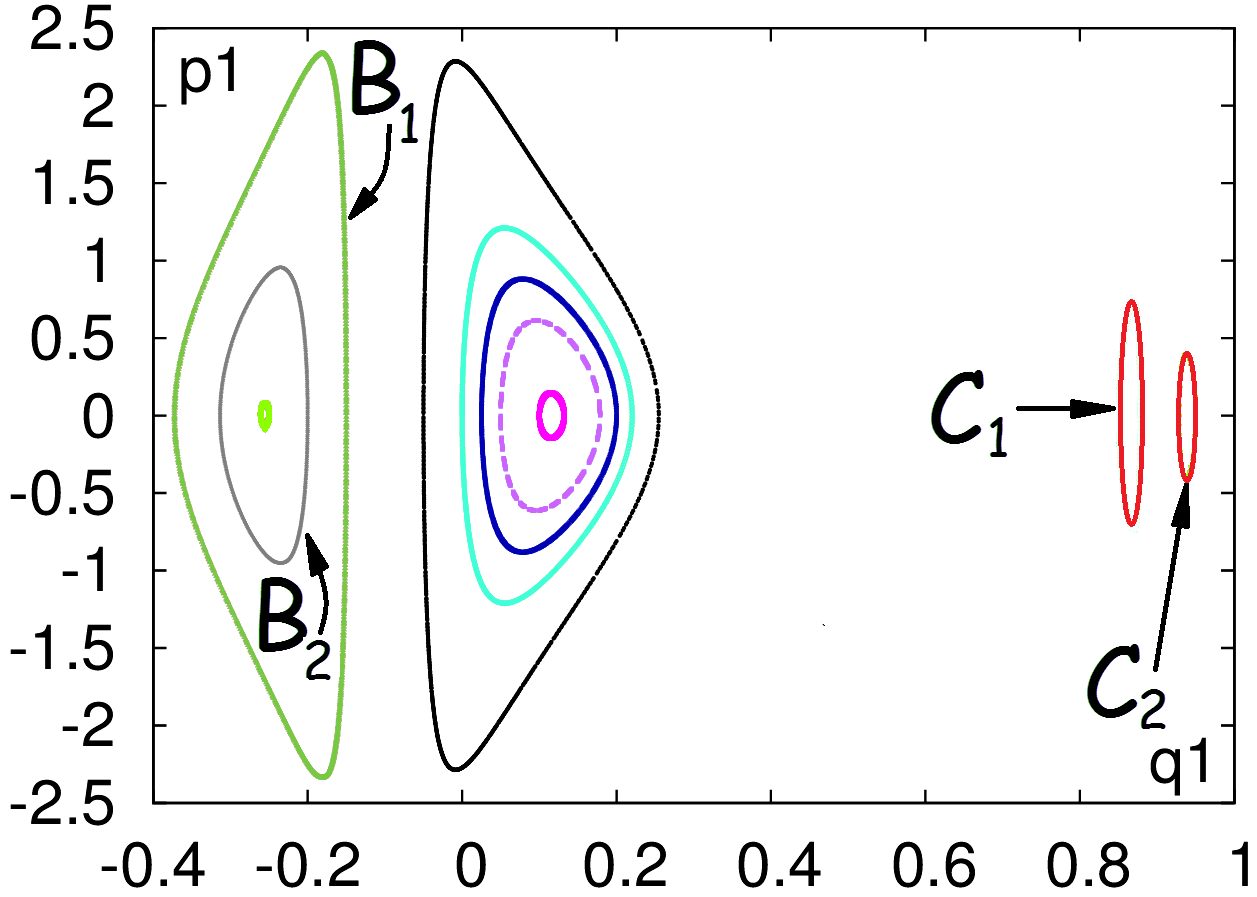}
\includegraphics[width=.235\textwidth]{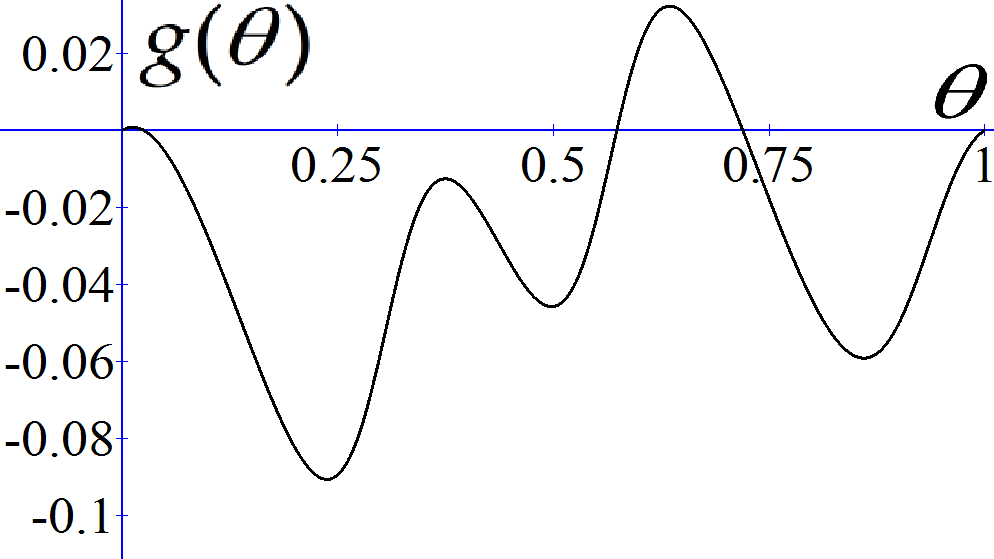}
\includegraphics[width=.235\textwidth]{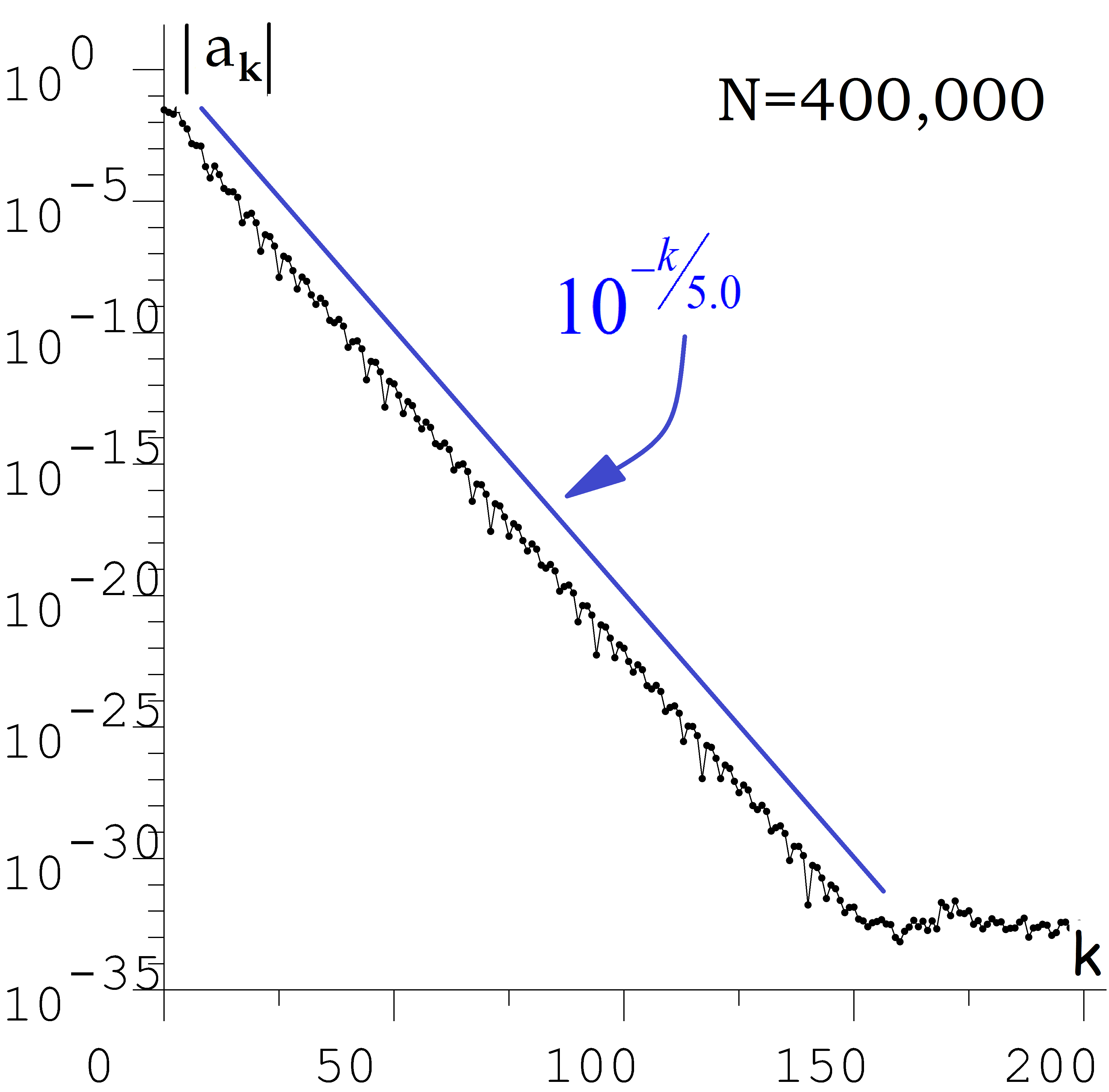}
\caption{\label{fig3bp} 
{\bf The restricted three-body problem.} Top left: A single 
quasiperiodic trajectory lying on a torus. The color indicates the value of the fourth variable $p_2$. Top right: Poincar{\'e} return map
for a variety of quasiperiodic trajectories, marked as $B_1, B_2, C_1, $ and $C_2$. The black curve in the top-left panel corresponds to $B_1$ and resides at $q_2 = 0$ where $dq_2/dt > 0$. 
Bottom left: 
The function $g(\theta) = \phi(\theta)-\theta$, the periodic part of the change of coordinates between the return map of 
orbit $B_1$ and pure rotation on a circle. Bottom right: The exponential decay of 
Fourier coefficients for orbit $B_1$, implying that $g$ is real analytic.
All orbits are for Hamiltonian $H =-2.63$. 
}
\end{figure}

%-_-_-_-_-_-_-_-_-_-_-_-_-_-_-_-_-_-_-_-_-_-_-_-_-_-_-_-_-_-_-_-_-_-_-_-_-_-_-_-_-_-_-_-_-_-_-_-_-_-_-_-_-_-_-_-_-_-_-_-_-_-_-_-_-_-_-_-_-_-_-_-_-_-_-_-
\section{The restricted three-body problem} 
Planetary motion is an application in which one would expect a high degree of quasiperiodicity. For example the moon's orbit has three rotation periods: 27.3 days, 8.85 years and 18.60 years (see \cite{moon_orbit}). It is quasiperiodic in rotating coordinates, filling out a three-dimensional torus in 6-dimensional phase space, when modeled as a circular restricted three-body problem in $\mathbb{R}^3$. (This ignores tides, other planets, and the eccentricity of the earth's orbit). We consider a planar three-body problem studied by Poincar{\'e}~\cite{ThreeBody2,ThreeBody1}. There are two massive bodies (``planet" and ``moon") moving in circles about their center of mass and a third body (``asteroid") whose mass is negligible, having no effect on the dynamics of the other two, all of which move in the same plane. The moon has mass $\mu = 0.1$ and the planet mass is $1-\mu$. We represent the bodies in rotating coordinates with the center of mass at $(0,0)$. The planet remains fixed at $(-0.1,0)$, and the moon is fixed at $(0.9,0)$. In these coordinates, the satellite's location and momentum are given by the {\em generalized position vector} $(q_1,q_2)$ and {\em generalized momentum vector} $(p_1,p_2)$. See~\cite{ThreeBody1, DSSY}
for the details of the equations of motion. The system's Hamiltonian $H(p_1,p_2,q_1,q_2)$, is the same for all orbits shown. Poincar{\'e} reduced this problem to the study of the Poincar{\'e} return map for a fixed value of $H$, only considering a discrete trajectory of the values of $(q_1,p_1)$ on the section $q_2=0$ and $dq_2/dt>0$. Thus we consider a
map in two dimensions rather than a flow in four dimensions. The top-left panel of Fig.~\ref{fig3bp} shows an orbit of the asteroid spiraling on a torus. The black curve is the
corresponding trajectory on the plane $q_2=0$, bordered in black.
\red{
%[[h2]]
 We used the order-8 Runge-Kutta method to compute the Poincar{\'e} section 
iterates of the three-body problem with time steps of $h=2\times 10^{-5}$}~\red{\cite{butcher08}.}
The top-right panel 
of Fig.~\ref{fig3bp} shows the Poincar{\'e} return map for the asteroid for a
variety of starting points, where orbit $B_1$ is the one shown in the top-left panel. 

Using $\Q_N$, we calculate the rotation number $\rho$ for orbit $B_1$, 
with 30-digit precision. As above we then use $\Q_N$ to 
calculate the Fourier coefficients for, now for  $B_1$.
See Fig.~\ref{fig3bp}, bottom-left. In the bottom-right panel of 
Fig.~\ref{fig3bp}, the 
Fourier coefficients converge exponentially fast, showing that the curve is real-analytic.\\

%-_-_-_-_-_-_-_-_-_-_-_-_-_-_-_-_-_-_-_-_-_-_-_-_-_-_-_-_-_-_-_-_-_-_-_-_-_-_-_-_-_-_-_-_-_-_-_-_-_-_-_-_-_-_-_-_-_-_-_-_-_-_-_-_-_-_-_-_-_-_-_-_-_-_-_-

\section{Discussion}
\red{
%[[c1]] 
The literature on quasiperiodicity is vast and our goals in this paper are limited: to introduce $\Q_N$ and the Embedding Continuation method for computing rotation numbers and to give some applications. Quasiperiodic orbits occur in a variety situations that we have not addressed.  For example Luque and Villanueva. \cite{luque:villanueva:14} made great progress with systems having external periodic forcing.  Medvedev et al.~\cite{medvedev_lagrangian_2015} consider high-dimensional tori that are not simply embedded. We also have not addressed noisy systems.}

%-_-_-_-_-_-_-_-_-_-_-_-_-_-_-_-_-_-_-_-_-_-_-_-_-_-_-_-_-_-_-_-_-_-_-_-_-_-_-_-_-_-_-_-_-_-_-_-_-_-_-_-_-_-_-_-_-_-_-_-_-_-_-_-_-_-_-_-_-_-_-_-_-_-_-_-
\acknowledgments 
\red{
We would like to thank the referee for many useful questions which helped us improve the manuscript. }%\red
CBD, MSF, and JW were  supported  by the NSF through the REU at the University of Maryland, College Park. 
ES was partially supported by NSF grant DMS-1407087. 
YS was partially supported by JSPS KAKENHI grant 26610034 and JST, CREST.
JY was partially supported by National Research Initiative Competitive grants 2009-35205-05209 and 2008-04049 from the USDA.

\bibliographystyle{unsrt_no_line}
\bibliography{Weighted_calc_bibliography_PhysicsStyle,1D_bibliography}

\end{document}